\UseRawInputEncoding

\documentclass[12pt]{article}

\usepackage{amssymb}
\usepackage{amsmath}
\usepackage{amsbsy} 

\usepackage{graphics}

\begin{document}


\title{\textbf{Can effective four-dimensional scalar theory be asymptotically
free in a spacetime with extra dimensions?}}

\author{
A.V. Kisselev\thanks{Electronic address: vladimir.petrov@ihep.ru}
{\small \ and}
V.A.~Petrov\thanks{Electronic address: alexandre.kisselev@ihep.ru} \\
{\small Division of Theoretical Physics, A.A. Logunov Institute for
High Energy Physics,}
\\
{\small NRC ``Kurchatov Institute'', 142281, Protvino, Russia}
}

\date{}

\maketitle

\begin{abstract}
We trace what happens with asymptotically free behavior of the
running coupling in $\phi^{3}$ theory in six-dimensional spacetime,
if to compactify two spatial dimensions on a 2D closed manifold. The
result can be considered as an effective 4D theory of infinitely
many KK-type scalar fields with triple interactions. The effective
\emph{dimensional} coupling constant inherits running to zero at
high mass scales in a modified form depending on the size of the
compact manifold. Some physical implications are discussed.
\end{abstract}


\hfill \parbox[r]{5.5 cm}{\emph{Freedom is not worth having if it
does not include the freedom to make mistakes.} \\
Mahatma Gandhi.}

\section{Introduction} %

Asymptotic freedom in QCD was discovered by David Gross and Frank
Wilczek \cite{Gross:1973}, and independently by David Politzer
\cite{Politzer:1973} in 1973. The asymptotically free renormalizable
field theory in four dimensions necessarily involves non-Abelian
gauge fields \cite{Coleman:1973, Bond:2017}. However, it is not the
case if a number of spacetime dimensions $D\neq 4$. The striking
examples are the 2D Gross-Neneu model \cite{Gross:1974} and 2D
nonlinear sigma model \cite{Brezin:1976}. All these theories are
renormalizable and asymptotically free. A special case is the 4D
$\phi^4$ theory with a negative coupling constant
\cite{Symanzik:1973}. It is a common belief that its spectrum can be
shown to be unbounded from below. Nevertheless, as was shown in
\cite{Brandt:1976}, this theory may be consistent. Especially note
the 6D scalar $\phi^3$ theory which also exhibits the property of
asymptotic freedom \cite{Macfarlane:1974}. One may ask is there any
4D \emph{effective} asymptotically free theory without gauge fields?
By effective theory we mean a reduced theory obtained from a higher
dimensional theory after ``integrating out'' extra spatial
coordinates. To answer our question, one has to consider theories in
a spacetime with extra dimensions (EDs).

Effective field theories with one or more compact EDs are of
considerable interest during last years. In particular, in
\cite{Elizade:1994} the total cross section of the scattering of two
light particle was calculated in the $\phi^4$ scalar model with a
spherical compactification. In \cite{Lopez:2020} one-loop order
contributions from one compact universal ED to the self-energy and
four-point vertex functions in a $\phi^4$ scalar theory are given.
The one-loop low-energy effective action in the $\phi^4$ scalar
theory and scalar QED with spacetime topology $\mathbb{R}^{3,1}
\otimes S^1$ is calculated in \cite{Akhoury:2009}. The decoupling of
heavy KK modes in Abelian Higgs model with spacetime topologies
$\mathbb{R}^{3,1} \otimes S^1$ and $\mathbb{R}^{3,1} \otimes
S^1/\mathbb{Z}_2$ is examined in \cite{Akhoury:2008}. The photon
self-energy, the fermion self-energy, and fermion vertex function in
the one-loop approximation in the context of QED with one ED are
presented in \cite{Martinez:2020}. In \cite{Sochichiu:1999} $D+1$
dimensional $\phi^3$ model with an arbitrary $D$ and one compact
manifold is studied. The renormalizable compactification models,
when a size of compact dimensions is of the order of cutoff scale,
are examined in \cite{Sochichiu:2000}. The universal extra
dimensional models defined on the six-dimensional spacetime with two
spatial dimensions compactified to a two-sphere orbifold $S^2/Z_2$
were studied in \cite{Dohi:2010}-\cite{Dohi:2014}. In
\cite{Kakuda:2013} $T^2/Z_2$, $S^2/Z_2$, and other orbifolds were
examined.

The goal of our study is to derive an effective four-dimensional
$\phi^3$ scalar field theory in a spacetime with two compact EDs and
calculate a running coupling constant in the one-loop approximation.
There are three possibilities to realize a scalar theory with a
power interaction $g\phi^n$ which has a \emph{dimensionless}
coupling constant $g$, see Tab.~\ref{tab:n_vs_D}. Among them only
the scalar $g\phi^3$ theory in six dimensions is known to be
asymptotically free \cite{Macfarlane:1974} (see also
\cite{Collins}). That is why, we will start from this theory.
%
\begin{table}
\begin{tabular}{||c||c|c|c||}
  \hline
  \textbf{D} & 3 & 4 & 6 \\
  \hline
  \textbf{n} & 6 & 4 & 3 \\
  \hline
\end{tabular}
\centering \caption{The dependence of an integer power $n$ on a
number of spacetime dimensions $D$ in scalar theories with an
interaction $g\phi^n(x)$ and \emph{dimensionless} coupling constant
$g$.} \label{tab:n_vs_D}
\end{table}

The paper is organized as follows. In Section~2 we briefly remind a
renormalization of the $\phi^3$ theory in \emph{six infinite}
dimensions (denoted hereafter as $\phi^3_6$, with the subscript 6
indicating the spacetime dimensionality). In the next section we
examine an effective $\phi^3$ theory in the spacetime with
\emph{four infinite and two compact} dimensions (referred below as
$\phi^3_{\mathrm{eff}}$) and calculate a running coupling constant.
In Section~4 we examine a dependence of our results on a topology of
the compact dimensions. Finally, in Section~5 a scale dependence of
physical observables is analyzed. Some properties of two-dimensional
inhomogeneous Epstein zeta function and truncated Epstein-like zeta
function are collected in Appendix~A.

\section{$\boldsymbol{\phi^3}$ theory in six infinite dimensions} %

The classical Lagrangian for the $\phi^3_6$ theory in terms of bare
parameters looks like
\begin{equation}\label{Lagrangian_4_dim_free}
\mathcal{L} =  \frac{1}{2} [\partial_\mu \phi(x)]^2 - \frac{1}{2}
m_0^2 \phi^2(x) - \frac{g_0}{3!} \phi^3(x) \;,
\end{equation}
where the bare coupling constant $g_0$ has a dimensionality of mass.
On a classical level a cubic potential of the $\phi^3$ theory is not
bounded below. As a consequence, there cannot be a stable ground
state. However, it is not the case, if one consider the theory on a
quantum level and takes into account a kinetic term in a
Hamiltonian, along with the cubic and quadratic ones
\cite{Kosyakov:2001}. In terms of the renormalized (R) field
$\phi_\mathrm{R}$, mass $m$ and coupling $g$ the Lagragian is given
by
\begin{equation}\label{Lagrangian_4_dim}
\mathcal{L} = \mathcal{L}_\mathrm{R} + \mathcal{L}_{\mathrm{CT}} \;,
\end{equation}
where
\begin{equation}\label{Lagrangian_4_renor}
\mathcal{L}_\mathrm{R} =  \frac{1}{2} [\partial_\mu
\phi_\mathrm{R}(x)]^2 - \frac{1}{2} m^2 \phi_\mathrm{R}^2(x) -
\frac{g}{3!} \phi_\mathrm{R}^3(x)
\end{equation}
is its renormalized part, and the counterterm part of
\eqref{Lagrangian_4_dim} is of the form
\begin{equation}\label{Lagrangian_4_renor}
\mathcal{L}_{\mathrm{CT}} =  \frac{1}{2} (Z_\phi - 1) [\partial_\mu
\phi_\mathrm{R}(x)]^2 - \frac{1}{2} \delta m^2 \phi_\mathrm{R}^2(x)
- (Z_\Gamma - 1)\frac{g}{3!} \phi_\mathrm{R}^3(x) \;.
\end{equation}
The Feynman rules are $i/(p^2 - m^2)$ for a scalar propagator, and
$(-ig)$ for a three-particle vertex. Let $\Gamma^{(n)}(p_1,p_2,
\ldots p_{n-1})$ be one-particle irreducible (OPI) Green's function.
The inverse propagator is given by
\begin{equation}\label{inverse_prop}
S^{-1}(p^2) = -i\,[p^2 - m^2 + \Sigma (p^2)] = -i\Gamma^{(2)}(p^2)
\;,
\end{equation}
where $\Sigma (p^2)$ is a self-energy. $\Gamma^{(3)}(p,q)$ is a
three-particle vertex with ``amputated'' external legs.

The renormalized quantities $(\phi_\mathrm{R}, g, m)$ are related
with the bare quantities $(\phi, g_0, m_0)$ through renormalization
constants (see, for instance, \cite{Collins}). In particular, the
scalar field is renormalized as
\begin{equation}\label{field_renorm_def}
\phi_\mathrm{R} = Z_{\phi}^{-1/2} \phi \;.
\end{equation}
The mass renormalization looks like
\begin{equation}\label{mass_renorm}
m^2 =  Z_m^{-1} m_0^2 \;.
\end{equation}
The renormalization of the coupling constant is given by
\begin{equation}\label{coupling_renorm}
g = Z_\phi^{3/2} Z_\Gamma^{-1} g_0 \;.
\end{equation}
If we express in \eqref{Lagrangian_4_dim} all the parameters in
terms of the bare quantities using
eqs.~\eqref{field_renorm_def}-\eqref{coupling_renorm}, we come to
\eqref{Lagrangian_4_dim_free}.

In our study, we use the dimensional regularization
\cite{t'Hooft:1973} for Feynman integrals, and the MOM scheme with
the Euclidean normalization point $-\mu^2$ ($\mu^2 > 0$) for the
renormalization procedure. Usually an on-shell condition is imposed
on propagators and vertices of scalar fields. In massive theories
where the zero momentum lies in the analyticity domain, a
subtraction point $p^2 = 0$ is used \cite{Collins}. Nevertheless, it
is more appropriate for us to normalize OPI Green's functions at
some Euclidean point, as it is done in QCD \cite{Gross:1973,
Yndurain}, where quarks and gluons are confined, and, consequently,
have no pole masses.

The beta function of the $\phi^3_6$ theory,
\begin{equation}\label{beta_fun}
\beta[g(\mu)] = \mu \frac{d g(\mu)}{d\mu} \;,
\end{equation}
is known to be \cite{Macfarlane:1974, Collins, Vasiliev}
\begin{equation}\label{beta_fun_expression}
\beta(g) = - \beta_0 g^3 + \mathrm{O}(g^5) \;,
\end{equation}
where
\begin{equation}\label{beta_0}
\beta_0 = \frac{3}{4(4\pi)^3} \;.
\end{equation}
It is calculated up to five loops \cite{Kompaniets:2021}. All known
terms in an expansion of $\beta(g)$ are negative. Since $\beta_0 >
0$, there is the \emph{asymptotic freedom} in $\phi_6^3$ theory, and
\begin{equation}\label{asymp_free}
\alpha(\mu) =  \frac{\alpha(\mu_0)}{\displaystyle { 1 + \frac{3}{4}
\alpha(\mu_0) \ln(\mu^2/\mu_0^2)}}  \;,
\end{equation}
where
\begin{equation}\label{alpha}
\alpha = \frac{g^2}{(4\pi)^3} \;.
\end{equation}
Note that, instead of using eq.~\eqref{beta_fun}, the
$\beta$-function can be alternatively defined as
\begin{equation}\label{beta_fun_alt}
\beta[g(\bar{\mu})] = - \bar{\mu} \frac{d g(\bar{\mu})}{d\bar{\mu}}
\;,
\end{equation}
where $\bar{\mu}$ is a scale needed to preserve the canonical
dimension of the coupling constant in the dimensional
regularization. The reason is that the renormalization constants
$Z_\phi$ and $Z_\Gamma$ depend on the ratio $\mu/\bar{\mu}$.

\section{$\boldsymbol{\phi^3}$ theory in spacetime with two extra compact dimensions} %

Let us consider $\phi^3$ theory in a spacetime with two extra
coordinates $y_1$, $y_2$, and metric tensor
\begin{equation}\label{metric_6_dim}
G_{MN} = (1, -1, -1, -1, \eta_{mn}) = (\gamma_{\mu\nu},\eta_{mn})
\;,
\end{equation}
where $M,N = (\mu, m)$, $\mu = 0, 1, 2, 3$, $m=1,2$, and $\eta_{mn}$
stands for the metric tensor of a 2D compact manifold. The scalar
field $\phi(x,y)$ is assumed to be defined on a manifold $M_4
\otimes T^2/Z_2$ with equal compactification radii $R_c$. Thus, the
field fulfills the periodicity and parity conditions
\begin{align}\label{y_i_interval}
\phi(x, y) &= \phi(x, y + 2\pi R_c) \;, \nonumber \\
\phi(x, y) &= \phi(x, - y) \;,
\end{align}
where $y = (y_1, y_2)$. A manifold with another topology will be
considered in Section~4.

The action in six dimensions with two compact dimensions is given by
the following expression
\begin{align}\label{action_4+2_dim}
\mathcal{S}_{4+[2]} = \int \!d^4 x \!\!\!\int\limits_{-\pi R_c}^{\pi
R_c} \!\!dy_1 \!\!\!\int\limits_{-\pi R_c}^{\pi R_c} \!\!dy_2 \,
\sqrt{-G} \, &\Big[ \frac{1}{2}
\partial_M \phi(x,y) \partial^M \!\phi(x,y) \nonumber  \\
&- \frac{1}{2} m^2 \phi^2(x,y) - \frac{g}{3!} \phi^3(x,y) \Big] ,
\end{align}
where $G = \mathrm{det}(G_{MN})$. The canonical dimension of
$\phi(x,y)$ is equal to 2. The coupling constant $g$ is
dimensionless. It is clear that in the limit $R_c \rightarrow
\infty$ the action \eqref{action_4+2_dim} becomes an 6D action of a
scalar field with interaction $g\phi^3$ in six infinite spacetime
dimensions (see the previous section).

We can use the following Fourier expansion of the field
\begin{equation}\label{phi_expansion}
\phi(x,y) = \frac{1}{2\pi R_c} \!\sum_{n_1 = -\infty}^{\infty}
\sum_{n_2 = -\infty}^{\infty} \!\!\!e^{i(n_1 y_1 + n_2 y_2)/R_c}
\phi_n(x) \;,
\end{equation}
where $n = (n_1,  n_2)$. Correspondingly, we have
\begin{equation}\label{phi_Fourier}
\phi_n(x) = \frac{1}{2\pi R_c} \!\!\int\limits_{-\pi R_c}^{\pi R_c}
\!\!dy_1 \!\!\!\int\limits_{-\pi R_c}^{\pi R_c} \!\!dy_2 \,e^{-i(n_1
y_1 + n_2 y_2)/R_c} \phi(x,y) \;.
\end{equation}
Note that every KK mode has canonical dimension 1.

If we require that the Kaluza-Klein (KK) modes $\phi_n(x)$ are
normalized,
\begin{equation}\label{pni_normalizaion}
\int \!\!d^4x \phi_{n}(x) \phi_{n'}(x) = \delta_{n,n'} \;,
\end{equation}
then
\begin{equation}\label{pni_normalizaion_y}
\int \!\!d^4x \!\!\int \!\!d^2y \,\phi_{n}(x,y)
\phi_{n'}^{\ast}(x,y) = \delta_{n,n'} \;.
\end{equation}
The masses of the KK excitations are
\begin{equation}\label{KK_masses}
m_n^2 = m_0^2 + \frac{n^2}{R_c^2} \;,
\end{equation}
where $n^2 = n_1^2 + n_2^2$, and $m_0$ means zero mode mass. Thus,
the \emph{effective} 4D action is given by
\begin{align}\label{action_effective}
\mathcal{S}_{4\mathrm{eff}} &= \int \!\!d^4 x \sqrt{-\gamma} \,
\Big\{ \frac{1}{2}
\partial_\mu \phi_0(x) \partial^\mu \phi_0(x) - \frac{1}{2} m_0^2 \phi_0^2(x)
\nonumber  \\
&-\sum_{n \neq 0} \, \Big[\,\frac{1}{2} \partial_\mu \phi_n(x)
\partial^\mu \phi_n(x) - \frac{1}{2} m_n^2 \phi_n^2(x)\, \Big] \nonumber  \\
&- \frac{g_4}{3!} \, \Big[ \phi_0^3(x) + \phi_0(x) \sum_{n \neq 0}
\, \phi_n(x) \phi_{-n}(x)
\nonumber  \\
&+ \sum_{n, m, k \neq 0} \!\! \phi_n(x) \phi_m(x) \phi_k(x) \,
\delta_{n+m+k, 0} \Big] \Big\} \;,
\end{align}
where $\gamma = \mathrm{det}(\gamma_{\mu\nu})$. Here
\begin{equation}\label{coupling_effective}
g_4 = \frac{g}{2\pi R_c}
\end{equation}
is an \emph{effective four-dimensional} coupling constant. Thus, it
is the inverse compactification scale $R_c^{-1}$ that makes $g_4$ a
quantity with the dimension of mass.

\subsection{Effective four-dimensional propagator in one-loop approximation} %

One of our main goals is a calculation of a scale dependence of the
coupling constant $g_4$ \eqref{coupling_effective}. As one can see
from \eqref{action_effective}, it is the same for zero mode
interactions, interactions between zero and KK modes, and non-zero
mode interactions. That is why, here and in what follows it is
assumed that all external particles have zero KK numbers. From the
very beginning, we put $m_0 = 0$.

\begin{figure}
\begin{center}
\resizebox{7cm}{!}{\includegraphics{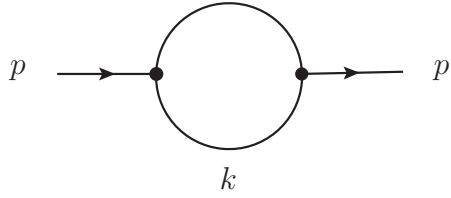}}
\caption{The self-energy diagram for the scalar field in the
$\phi^3$ theory in the one-loop approximation.}
\label{fig:prop}
\end{center}
\end{figure}
The four-dimensional self-energy of the scalar field at order
$\mathrm{O}(g^2)$ is given by the diagram in Fig.~\ref{fig:prop}. It
can be divided into two parts
\begin{equation}\label{Sigma_expansion}
\Sigma (p^2) = \Sigma_0(p^2) + \Sigma_{\mathrm{KK}}(p^2) \;,
\end{equation}
where
\begin{equation}\label{Sigma_0}
\Sigma_0(p^2) = -\frac{i}{2} \,g_4^2 \, \bar{\mu}^{2\epsilon}
\int\limits_0^1 \!\!dx \!\!\int \!\!\frac{d^D\!k}{(2\pi)^D} \,
\frac{1}{[k^2 + p^2x(1-x)]^2}
\end{equation}
is the contribution from zero mode, and
\begin{equation}\label{propagator_n}
\Sigma_{\mathrm{KK}}(p^2) = -\frac{i}{2} \, g_4^2 \,
\bar{\mu}^{2\epsilon} \sum_{n \neq 0}\int\limits_0^1 \!\!dx \!\!\int
\!\!\frac{d^D\!k}{(2\pi)^D} \, \frac{1}{[k^2 + p^2x(1-x) - m_n^2]^2}
\end{equation}
is the contribution from KK massive modes. It is assumed that $p^2 <
0$. We define
\begin{equation}\label{4_dim}
D = 4 - 2\varepsilon \;.
\end{equation}
We find that
\begin{align}\label{Sigma_0}
\Sigma_0(p^2) &= \frac{g_4^2}{2(4\pi)^{2 -\epsilon}} \,
\Gamma(\epsilon) \left( \frac{\bar{\mu}^2}{-p^2} \right)^{\!
\varepsilon} \int\limits_0^1 \!\!dx [x(1-x)]^{-\varepsilon}
\nonumber \\
&= \frac{\alpha}{2\pi}  R_c^{-2} \left[ N_\varepsilon - \ln
\frac{-p^2}{\bar{\mu}^2} + 2 \right] + \mathrm{O}(\varepsilon) \;,
\end{align}
where
\begin{equation}\label{N_eps}
N_\varepsilon = \frac{1}{\varepsilon} - \gamma_E + \ln 4\pi \;.
\end{equation}
Thus, zero mode contributes to the mass renormalization only.

Now we consider the contribution from the massive modes
\begin{align}\label{Sigma_KK}
\Sigma_{\mathrm{KK}}(p^2) &= -\frac{i}{2} \,g_4^2\, (\bar{\mu}
R_c)^{2\epsilon} R_c^{-2} \,\sum_{n_1, n_2 \neq 0} \int\limits_0^1
\!\!dx \!\!\int \!\!\frac{d^D\!l}{(2\pi)^D} \nonumber \\
&\times \frac{1}{[l^2 + p^2R_c^2 x(1-x)  - n_1^2 - n_2^2]^2} \;,
\end{align}
where $l = k R_c$. Since
\begin{align}\label{Sigma_KK_int}
&\int \!\!\frac{d^D\!l}{(2\pi)^D} \frac{1}{[l^2 + p^2R_c^2 x(1-x) -
n_1^2 - n_2^2]^2} \nonumber \\
&= \frac{i}{(4\pi)^{2-\varepsilon}} \,\Gamma(\varepsilon) [-p^2
x(1-x) + n_1^2 + n_2^2]^{-\varepsilon} \;,
\end{align}
we obtain
\begin{align}\label{Sigma_KK_2}
\Sigma_{\mathrm{KK}}(p^2) &= \frac{\alpha}{2\pi}
\,\Gamma(\varepsilon) (4\pi)^\varepsilon (\bar{\mu} R_c)^{2\epsilon}
R_c^{-2} \nonumber \\
&\times \!\!\sum_{n_1, n_2 \neq 0} \int\limits_0^1 \!\!dx [-p^2R_c^2
x(1-x) + n_1^2 + n_2^2]^{-\varepsilon} \;.
\end{align}
The series in \eqref{Sigma_KK_2} converges absolutely for
$\mathrm{Re} \,\varepsilon > 1$. To define this series for other
values of $\varepsilon$, we require its analytic continuation using
two-dimensional inhomogeneous Epstein zeta function $Z^{a}_2(s)$
\cite{Kirsten:1991}
\begin{equation}\label{Epstein_func_def}
Z^{a}_2(s) = \sum_{n_1, n_2 \in Z^{\!2}}{}^{\!\!\!\!\!'}
\frac{1}{(n_1^2 + n_2^2 + a)^s} \;,
\end{equation}
with $a>0$ (the prime in \eqref{Epstein_func_def} means that the
point $n = 0$ is to be excluded from the sum). The zeta function
regularization method for the quantum physical systems was proposed
for the first time in \cite{Dowker:1976, Hawking:1997}. The Riemann
zeta function $\zeta(s)$ was used in fixing a critical spacetime
dimension of the string theory (see, for instance, \cite{Zwiebach}).
Recently, one-dimensional inhomogeneous Epstein zeta function
$Z^{a}_1(s)$ was applied to quantify the UV divergences induced by
the KK fields \cite{Lopez:2020}-\cite{Akhoury:2008}. In
\cite{Martinez:2020} both $Z^{a}_1(s)$ and $n$-dimensional
inhomogeneous function $Z^{a}_n(s)$ were used.

In Appendix~A formula \eqref{Epstein_full} is presented which gives
an analytical continuation for the function $Z^{a}_2(s)$. It is
defined on the complex plane of $s$. It has an infinite number of
simple poles, but \emph{converges} both in the $s \rightarrow 0$
limit, and with $a =0$. These results is a consequence of the
analytical properties of the inhomogeneous Epstein zeta function.

Let us define
\begin{equation}\label{c}
c = -p^2 R_c^2 x(1-x) \;.
\end{equation}
Note that $c>0$, except for two points $x=0, 1$. We obtain from
\eqref{Sigma_KK_2}-\eqref{c}
\begin{equation}\label{Sigma_KK_3}
\Sigma_{\mathrm{KK}}(p^2) = \frac{\alpha}{2\pi}
\,\Gamma(\varepsilon) (4\pi)^\varepsilon (\bar{\mu} R_c)^{2\epsilon}
R_c^{-2} \!\int\limits_0^1 \!\!dx Z^{c}_2(\varepsilon) \;,
\end{equation}
where
\begin{equation}\label{Epstein_func_m}
Z^{c}_2(\varepsilon) = - c^{-\varepsilon} - \frac{\pi c^{1 -
\varepsilon}}{1 - \varepsilon} +
\frac{A(\varepsilon;c)}{\Gamma(\varepsilon)} \;.
\end{equation}
Since $Z^{c}_2(a)$ is finite for $c=0$, see \eqref{Epstein_zero_a},
we can take $c>0$ for all $x \in [0,1]$. An explicit expression for
$A(s;a)$ in \eqref{Epstein_func_m} is given by eq.~\eqref{A}. The
function $A(\varepsilon;c)$ converges, as $\varepsilon \rightarrow
0$, and, consequently, $Z^{c}_2(0) = -(1 + \pi c)$. The KK
divergence (the first term in \eqref{Epstein_func_m}) exactly
cancels the zero mode divergence \eqref{Sigma_0}. A similar effect
was seen in the context of quantum electro\-dynamics with one ED
\cite{Akhoury:2009}. Since $A(\varepsilon;c)$ decreases
exponentially as $c \rightarrow \infty$, we find for large $R_c$
\begin{align}\label{Sigma_tot}
\Sigma(p^2) &= p^2 \frac{\alpha}{2} \Gamma(\varepsilon)
(4\pi)^\varepsilon \left( \frac{\bar{\mu}^2}{-p^2} \right)^{\!
\varepsilon} \int\limits_0^1
\!\!dx [x(1-x)]^{1-\varepsilon} \nonumber \\
&= p^2 \,\frac{\alpha}{12} \left( N_\varepsilon - \ln
\frac{-p^2}{\bar{\mu}^2}+ \frac{5}{3} \right) +
\mathrm{O}(\varepsilon) \;.
\end{align}
As a result, for $\mu R_c \gg 1$, the field renormalization constant
is equal to
\begin{equation}\label{Z_phi_4_eff}
Z_{\phi} = 1 - \frac{\alpha}{12} \left( N_\varepsilon - \ln
\frac{\mu^2}{\bar{\mu}^2} + \frac{5}{3} \right) .
\end{equation}
It differs from the field renormalization in the $\phi^3_6$ theory
by a constant term only. Note, there is no dependence on $R_c$ in
\eqref{Z_phi_4_eff}. Since $\Sigma(p^2) \sim p^2$, the renormalized
theory remains massless in the one-loop approximation (no mass
renormalization holds).

\subsection{Effective four-dimensional vertex in one-loop approximation} %

\begin{figure}
\begin{center}
\resizebox{6cm}{!}{\includegraphics{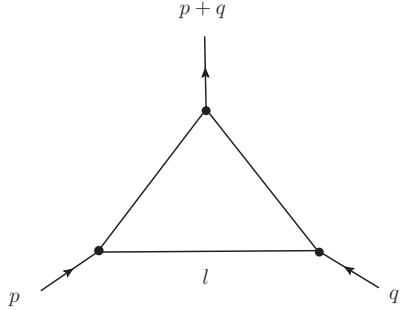}}
\caption{Three-particle vertex in the scalar $\phi^3$ theory in the
one-loop approximation.} \label{fig:vertex}
\end{center}
\end{figure}

The effective four-dimensional three-point vertex $\Gamma^{(3)}$ is
defined by the diagram presented in Fig.~\ref{fig:vertex}. It is a
sum of two terms,
\begin{equation}\label{Gamma_expansion}
\Gamma^{(3)}(p,q) = \Gamma_0^{(3)}(p,q) +
\Gamma_{\mathrm{KK}}^{(3)}(p,q) \;,
\end{equation}
where
\begin{equation}\label{Gamma_0}
\Gamma_0^{(3)}(p,q) = 2g_4^3 \bar{\mu}^{2\epsilon}
\!\!\int\limits_0^1 \!dx \,x \!\!\int\limits_0^1 \!dy \!\!\int
\!\!\frac{d^D k}{(2\pi)^D} \, \frac{1}{(k^2 - M^2)^3} \;,
\end{equation}
and
\begin{align}\label{Gamma_KK}
\Gamma_{\mathrm{KK}}^{(3)}(p,q) &= 2g_4^3 \bar{\mu}^{2\epsilon}
\sum_{n_1, n_2 \neq 0} \int\limits_0^1 \!dx \,x \!\!\int\limits_0^1
\!dy \!\!\int \!\!\frac{d^D k}{(2\pi)^D} \, \frac{1}{(k^2 - M^2 -
m_n^2)^3} = 2g_4^3 \bar{\mu}^{2\epsilon} R_c^{2+2\varepsilon} \nonumber \\
&\times \sum_{n_1, n_2 \neq 0} \int\limits_0^1 \!dx \, x
\!\!\int\limits_0^1 \!dy \!\!\int \!\!\frac{d^D l}{(2\pi)^D} \,
\frac{1}{(l^2 - M^2R_c^2 - n_1^2 - n_2^2)^3} \;.
\end{align}
Here a notation
\begin{equation}\label{M2}
M^2 =-x[ p^2xy(1-y) + q^2y(1-x) + (p+q)^2(1-x)(1-y) ]
\end{equation}
is introduced. We assume that $p^2, q^2, (p+q)^2 < 0$. It means that
$M^2 > 0$, except for two points $(x,y) = (1,0), (1,1)$. The
integral in \eqref{Gamma_0} converges, as $\varepsilon \rightarrow
0$, and we obtain
\begin{equation}\label{Gamma_0_2}
\Gamma_0^{(3)}(p,q) = (-ig_4)\frac{\alpha}{\pi} \,
\Gamma(1+\varepsilon) (4\pi)^\varepsilon \bar{\mu}^{2\epsilon}
R_c^{-2} \!\!\int\limits_0^1 \!dx \,x \!\!\int\limits_0^1 \!dy
(M^2)^{-1 - \varepsilon} \;.
\end{equation}
In particular, we find for $\varepsilon = 0$
\begin{equation}\label{Gamma_0_norm}
\Gamma_0^{(3)}(p,q)|_{p^2 = q^2 = (p+q)^2 = - \mu^2} =
(-ig_4)\frac{\alpha}{2\pi} \,B(\mu R_c)^{-2} \;,
\end{equation}
where
\begin{align}\label{B}
B &= 2\int\limits_0^1 \!dx \!\!\int\limits_0^1 \!dy \,[1-x +
xy(1-y)]^{-1} \nonumber \\
&= \frac{1}{27} \left[ \psi_1 \!\!\left( \frac{1}{3} \right) +
\psi_1 \!\!\left( \frac{1}{6} \right) - \psi_1 \!\!\left(
\frac{5}{6} \right)  - \psi_1\!\!\left( \frac{2}{3} \right) \right]
,
\end{align}
$\psi_1(z) = (d^2\!/dz^2)\ln \Gamma(z)$ being the trigamma function
\cite{Bateman_vol_1}.

The integral on the right-hand side of eq.~\eqref{Gamma_KK} is equal
to
\begin{align}\label{Gamma_KK_int}
&\int \!\!\frac{d^Dl}{(2\pi)^D} \, \frac{1}{(l^2 - M^2 R_c^2 - n_1^2
- n_2^2)^3} \nonumber \\
&= -\frac{i}{2(4\pi)^{2-\varepsilon}} \,\Gamma(1 + \varepsilon)
(M^2R_c^2 + n_1^2 + n_2^2)^{-1 - \varepsilon}  \;,
\end{align}
that results in
\begin{equation}\label{Gamma_KK_Epstein}
\Gamma_{\mathrm{KK}}^{(3)}(p,q) = (-ig_4)\frac{\alpha}{\pi} \,
\Gamma(1+\varepsilon) (4\pi)^\varepsilon (\bar{\mu} R_c)^{2\epsilon}
\int\limits_0^1 \!dx \, x \!\!\int\limits_0^1 \!dy
Z_2^{M^2\!R_c^2}(1+\varepsilon) \;.
\end{equation}
Thus, the infinite number of UV divergences results in the
two-dimensional inhomogeneous Epstein zeta function. We find from
eqs.~\eqref{Epstein_full}, \eqref{A}
\begin{equation}\label{Epstein_func_m_2}
Z^{M^2\!R_c^2}_2(1+\varepsilon) = - (M^2\!R_c^2)^{-1 - \varepsilon}
+ \frac{\pi (M^2\!R_c^2)^{-\varepsilon}}{\varepsilon} +
\frac{A(1+\varepsilon; M^2\!R_c^2)}{\Gamma(1+\varepsilon)} \;,
\end{equation}
with $A(1;c)$ being a finite quantity. As one can see from
\eqref{Epstein_func_m_2}, $Z^{M^2\!R_c^2}_2(1+\varepsilon)$ has a
simple pole at $\varepsilon = 0$. It can be easily shown that in the
limit $\varepsilon \rightarrow 0$ the fist term in
\eqref{Epstein_func_m_2}, after substitution in
\eqref{Gamma_KK_Epstein}, cancels the zero mode contribution
\eqref{Gamma_0_2}, and we obtain
\begin{align}\label{Gamma_tot}
\Gamma^{(3)}(p,q) &= (-ig_4)\frac{\alpha}{\pi} \,
\Gamma(1+\varepsilon) (4\pi)^\varepsilon (\bar{\mu} R_c)^{2\epsilon}
\nonumber
\\
&\times \int\limits_0^1 \!dx \, x \!\!\int\limits_0^1 \!dy \Bigg[
\frac{\pi (M^2\!R_c^2)^{-\varepsilon}}{\varepsilon} +
\frac{A(1+\varepsilon; M^2\!R_c^2)}{\Gamma(1+\varepsilon)} \Bigg] .
\end{align}
Thus, for $\varepsilon \rightarrow 0$ the vertex renormalization
constant is given by the expression
\begin{equation}\label{Z_gamma_4_eff}
Z_\Gamma = 1 - \frac{\alpha}{\pi} \int\limits_0^1 \!dx \,x
\!\!\int\limits_0^1 \!dy \Bigg[ \Gamma(\varepsilon)
\,\pi(4\pi)^{\varepsilon}\left( \frac{\bar{\mu}^2}{M^2_\mu}
\right)^{\!\!\varepsilon} + A(1; M^2_\mu\!R_c^2) \Bigg] .
\end{equation}
where
\begin{equation}\label{M2_mu}
M^2_\mu = \mu^2x[ 1-x + xy(1-y)] \;,
\end{equation}
$-\mu^2$ being the renormalization point. As one can see from
\eqref{Z_gamma_4_eff}, the vertex renormalization constant
$Z_\Gamma^{-1}$ depends both on the ratio $\mu/\bar{\mu}$ and on the
compactification radius via dimensionless parameter $\mu R_c$. The
vertex has a divergence related with a summation over KK number,
while Feynman integral is finite.

However, for $\mu R_c \gg 1$ (and, consequently, for $M^2_\mu\!R_c^2
\gg 1$), the function $A(1; M^2_\mu\!R_c^2)$ decreases exponentially
(see eq.~\eqref{A}), and we obtain
\begin{equation}\label{Z_gamma_4_eff_large_R}
Z_\Gamma^{-1} = 1 + \frac{\alpha}{2} \left[ N_\varepsilon - \ln
\!\left( \frac{\mu^2}{\bar{\mu}^2} \right)
 - C \right] ,
\end{equation}
where
\begin{equation}\label{C}
C = 2\int\limits_0^1 \!x dx \!\!\int\limits_0^1 \!dy \, \{ \ln x +
\ln [1-x + xy(1-y)] \} = \frac{2B}{3} - 3 \;.
\end{equation}
As we can see, if the compactification radius exceeds the physical
scale, $R_c \gg \mu^{-1}$, it disappears from the renormalization
constants \eqref{Z_phi_4_eff} and \eqref{Z_gamma_4_eff}.

The renormalized effective four-dimensional vertex is proportional
to $g_4$ \eqref{coupling_effective}. The fact that the coupling of
the four-dimensional fields becomes smaller at larger $R_c$ can be
easily understood. As it follows from \eqref{phi_expansion}, the
wave function of the field $\phi_n(x)$ in the $y$-space is given by
\begin{equation}\label{wave_fun}
\psi_n (y) = \frac{1}{2\pi R_c} \, e^{i n y/R_c} \;.
\end{equation}
The coupling constant of three fields $\phi_n (x)$, $\phi_m (x)$,
$\phi_k(x)$ is defined by overlapping of their wave functions
\begin{equation}\label{3_fields_coupl}
g \!\!\!\int\limits_{-\pi R_c}^{\pi R_c} \!\!dy_1
\!\!\int\limits_{-\pi R_c}^{\pi R_c} \!\!dy_2 \, \psi_n (y) \psi_m
(y) \psi_k (y) = \frac{g}{2\pi R_c} \,\delta_{n+m+k, 0} \;.
\end{equation}
It tends to zero as $R_c$ grows. Thus, in the limit $R_c \rightarrow
\infty$ (all six dimensions are infinite), the
$\phi^3_{\mathrm{eff}}$ theory becomes a theory of a \emph{free}
scalar field, whose propagator is equal to that of the $\phi^3_4$
theory.

\subsection{Running coupling constant} %

Let us consider large values of the mass scale $\mu$, namely, $\mu
\gg R_c^{-1}$. It follows from eqs.~\eqref{coupling_renorm},
\eqref{Z_phi_4_eff} and \eqref{Z_gamma_4_eff_large_R} that in the
one-loop approximation the beta function is equal to
\begin{equation}\label{beta_fun_4}
\beta(g) = -\frac{3R_c^2}{64\pi} \,g^3 \;,
\end{equation}
and, correspondingly,
\begin{equation}\label{alpha_4_mu_dependence}
\mu^2 \frac{\partial \alpha_4(\mu)}{\partial \mu^2} =
-\frac{3R_c^2}{16} \,\alpha_4^2(\mu) \;,
\end{equation}
where
\begin{equation}\label{alpha_4}
\alpha_4 = \frac{g_4^2}{4\pi} \;.
\end{equation}
Let us note, it is the \emph{dimensional} variable $R_c^2\ln
(\mu^2/\mu_0^2)$, not the dimensionless quantity $\ln
(\mu^2/\mu_0^2)$, which should be regarded as an evolution parameter
for the coupling constant $\alpha_4(\mu)$. It is to be expected,
since the coupling $\alpha_4$ has dimension $-2$. As a result, we
obtain
\begin{equation}\label{asymp_freedom}
\alpha_4(\mu) = \frac{\alpha_4(\mu_0)}{\displaystyle { 1 +
\frac{3}{16} \alpha_4(\mu_0) R_c^2 \ln(\mu^2/\mu_0^2)}} =
\frac{16}{3 R_c^2 \ln(\mu^2\!/\Lambda^2) }  \;,
\end{equation}
where
\begin{equation}\label{Lambda}
\Lambda^2 = \mu_0^2 \exp\big[-16/(3 \alpha_4(\mu_0) R_c^2) \big] =
\mu_0^2 \exp\!\big[-4/(3 \alpha(\mu_0) \big] .
\end{equation}
We remind that eqs.~\eqref{asymp_freedom} and \eqref{Lambda} hold in
the one-loop approximation and at $\mu \gg \Lambda$. Ghost pole at
$\mu = \Lambda$ is safely eliminated if to respect the causality
\cite{Shirkov:1997}.

Thus, the effective four-dimensional scalar $\phi^3$ theory in the
flat spacetime with two compact EDs exhibits \emph{the property of
asymptotic freedom}. Namely, its effective coupling constant
$\alpha_4(\mu)$ tends logarithmically to zero, as the mass scale
$\mu$ grows. One can say that four-dimensional theory does not
forget its higher dimensional origin.

All this can be understood as follows. The renormalization of the
coupling constant is defined by the UV divergences and
renormalization scale $\mu$, and ``it is not aware'' of the scale
$R_c^{-1}$, provided $\mu \gg R_c^{-1}$. In other words, the large
scale $R_c$ is irrelevant to a small-distance physics. As a result,
the effective four-dimensional coupling constant $g_4$ exhibits a
large-scale behavior of the coupling constant in the $\phi^3_6$
theory. For a detailed discussion of this phenomenon, see Section~5.

It is interesting to compare our prediction \eqref{beta_fun_4} with
the results obtained for an effective 4D $\lambda \phi^4$ theory in
a spacetime with one compact ED \cite{Akhoury:2009}. Is has been
found that in such a theory an effective coupling constant in
one-loop approximation is renormalized by the constant
\begin{equation}\label{coup_ren_phi4_4+1}
Z_\phi^{3/2} Z_\Gamma^{-1} = 1 + \frac{3\lambda^2}{16\pi^2} \left[
\frac{1}{\varepsilon } + \ln(\mu R_c) \right] .
\end{equation}
Note that $\lambda = \bar{\lambda}/(2\pi R_c)$, where
$\bar{\lambda}$ is the coupling constant in a 5D action with
dimension $-1$. Thus, one can not obtain a RG-like equation for
$\lambda$ with respect to the scale $\bar{M}=R_c^{-1}$, as it is
erroneously stated in \cite{Akhoury:2009} (see also
\cite{Sochichiu:1999}), except when $\bar{\lambda} =
\bar{\lambda}(R_c) = \mathrm{constant} \times R_c$. For instance, if
we assume that this relation takes place for small $R_c$, then we
come to the equation with respect to the intrinsic scale of the
spacetime topology,
\begin{equation}\label{RG_4+1_large_M_bar}
\frac{d \lambda}{d \ln \bar{M}} = -\frac{3\lambda^2}{16\pi^2} \;,
\end{equation}
valid for large $\bar{M}$.

As for the case $\mu \ll R_c^{-1}$, it can be shown that
$\alpha_4(\mu)$ tends to a constant value, as $\mu$ grows (while
being less than $R_c^{-1}$). As one see from \eqref{Sigma_KK_3}, the
total divergence in $\varepsilon = (4-D)/2$ is due to the UV
divergent Feynman integral, while the sum in the KK modes gives a
finite result ($Z_2^c(\varepsilon)$ is finite, as $\varepsilon
\rightarrow 0$). On the contrary, as eq.~\eqref{Gamma_KK_Epstein}
shows, the vertex divergence comes from the infinite sum over KK
modes only ($Z_2^c(1+\varepsilon) \sim \varepsilon^{-1}$, as
$\varepsilon \rightarrow 0$), while the Feynman integral is finite.
If $\mu \gg R_c^{-1}$, infinite and $\mu$-dependent parts of the
counterterms of the origin, six-dimensional, theory and those of the
reduced theory coincide. However, our calculations have shown that
$\mu$-dependent parts of the renormalization constants differ for
$\mu \ll R_c^{-1}$, and a nontrivial dependence on $R_c$ occurs. Let
us note that the divergent $\varepsilon^{-1}$ terms remain the same
regardless of a value of $R_c$, in a full accordance with the
results of \cite{Duff:1981}. It is to be expected, since the
compactification is an infrared process which can not change the UV
properties of the theory.

\section{Compactification on orbifold $S^2/Z_2$} %

In Section~3 the manifold $M_4 \otimes T^2/Z_2$ was studied. In this
section we examine the case when the six-dimensional scalar field
$\phi$ is defined on a manifold $M_4 \otimes S^2/Z_2$, with a radius
of two-dimensional sphere $S^2$ to be $R_c$. It is appropriate to
introduce spherical coordinates $\theta, \phi$, and use the
following expansion
\begin{equation}\label{phi_expan}
\phi(x,\theta, \phi) = \frac{1}{R_c} \!\sum_{l = 0}^{\infty} \sum_{m
= -l}^{l} Y_l^m(\theta, \phi) \,\phi_{lm}(x) \;,
\end{equation}
where $Y_l^m(\theta, \phi)$ ($m = -l, -l+1, \ldots, l-1, l$) are
spherical harmonics \cite{MacRobert}. They obey the orthogonality
condition
\begin{equation}\label{Y_normalization}
\int\limits_0^{2\pi} d\phi \int\limits_0^{\pi} \sin\theta d\theta
\,Y_l^m(\theta, \phi) [Y_{l'}^{m'}(\theta, \phi)]^* = \delta_{l l'}
\,\delta_{m m'} \;.
\end{equation}
Using formula
\begin{equation}\label{int_3_harmonics}
\int\limits_0^{2\pi} d\phi \int\limits_0^{\pi} \sin\theta d\theta
\,Y_0^0(\theta, \phi) Y_l^m(\theta, \phi) Y_l^{-m}(\theta, \phi) =
\frac{(-1)^m}{\sqrt{4\pi}} \;,
\end{equation}
one can show that an effective four-dimensional coupling constant is
\begin{equation}\label{coupling_eff_zero_modes}
\bar{g}_4 = \frac{g}{\sqrt{4\pi} R_c} \;,
\end{equation}
for zero mode interaction. For interactions between zero mode and KK
modes, a coupling constant is equal to $(-1)^m\bar{g}_4$. The masses
of the KK excitations are known to be numerated by an integer $l =
0, 1, 2, \ldots$ \cite{Dohi:2010,Maru:2010},
\begin{equation}\label{KK_massses_l}
m_l^2 = m_0^2 + \frac{l(l+1)}{R_c^2} \;.
\end{equation}

Let us consider zero-mode self-energy $\Sigma(p^2)$ in the one-loop
approximation (Fig.~\ref{fig:prop}). It is given by
\begin{align}\label{Sigma_KK_l}
\Sigma(p^2) &= \frac{\alpha}{2} \,\Gamma(\varepsilon)
(4\pi)^\varepsilon (\bar{\mu} R_c)^{2\epsilon} R_c^{-2}
\sum_{l=0}^\infty \sum_{m=-l}^l \int\limits_0^1 \!\!dx [l(l+1) +
c]^{-\varepsilon}
\nonumber \\
&= \frac{\alpha}{2} \,\Gamma(\varepsilon) (4\pi)^\varepsilon
(\bar{\mu} R_c)^{2\epsilon} R_c^{-2} \int\limits_0^1 \!\!dx
\,\sum_{l=0}^\infty \,\frac{2l+1}{[l(l+1) + c]^\varepsilon} \;,
\end{align}
where $c$ is defined by eq.~\eqref{c}. The series on the right-hand
side can be represented as
\begin{equation}\label{zeta_s_c}
\zeta_t(s;c) = \sum_{l=0}^\infty \frac{2l+1}{ [ l(l+1) + c]^s } =
\frac{1}{1-s} \,\frac{d}{d\alpha} \sum_{l=0}^\infty \frac{1}{ [
l(l+1) + \alpha(2l+1) + c ]^{s-1} } \,\Big|_{\alpha = 0} \;.
\end{equation}
We have
\begin{equation}\label{szeta_with_alpha}
\sum_{l=0}^\infty \frac{1}{ [ l(l+1) + \alpha(2l+1) + c ]^{s-1} } =
\sum_{l=0}^\infty \frac{1}{ [ (l+a)^2 + q ]^{s-1} } = \zeta_t(s;a,q)
\;,
\end{equation}
where
\begin{equation}\label{a_q}
a = \frac{1}{2} + \alpha \;, \quad q = c - \frac{1}{4} - \alpha^2
\;,
\end{equation}
and an analytic expression for $\zeta_t(s;a,q)$ is given by
eq.~\eqref{trancated_zeta_fun}. Note that $[dq/d\alpha]|_{\alpha=0}
= 0$. For $c \gg 1$, we obtain form \eqref{zeta_s_c}-\eqref{a_q},
and \eqref{trancated_zeta_fun} that
\begin{equation}\label{s_epsilon}
\zeta_t(\varepsilon;c) =  - c^{1-\varepsilon} \left[ 1 -
\frac{1}{12c} \right] + \mathrm{O}(c^{-2})  \;,
\end{equation}
as $\varepsilon \rightarrow 0$. As a result, we come to expression
\eqref{Z_phi_4_eff} (up to unimportant finite constant).

The above consideration can be also applied to a calculation of the
effective four-dimensional vertex for zero mode interaction in the
one-loop approximation (Fig.~\ref{fig:vertex}). Taking into account
that
\begin{equation}\label{s_1+epsilon}
\zeta_t(1+\varepsilon;c) =  c^{-\varepsilon} \left[
\frac{1}{\varepsilon} + \frac{1}{12c} \right] + \mathrm{O}(c^{-2})
\;,
\end{equation}
as $\varepsilon \rightarrow 0$, we reproduce formula
\eqref{Z_gamma_4_eff_large_R} (up to a constant factor). All said
above allows us to conclude that in the large $R_c$ region our main
results do not depend on a topology of the two-dimensional compact
manifold.

\section{Scale dependence of physical observables} %

\hfill \parbox[r]{8.1cm}{ \emph{Ultimately the circumference of an
infinite circle and a straight line are the same thing.} \\
Galileo Galilei, ``Dialogue concerning the two chief world systems:
Ptolemaic and Copernican. The third day.''}
\\ \\

As we already mentioned in Section~3, a nontrivial dependence of
physical quantities on the compactification radius appears when the
physical scale ($\mu^{-1}$, in our case) becomes much larger than
$R_c$. In the opposite case, $\mu^{-1} \lesssim R_c$, when a
physical process goes ``inside a sphere of the radius $R_c$'', such
a dependence disappears.

Some other physical examples can be given which illustrates these
statements. In \cite{Petrov:2001} a generalization of the
Froissart-Martin bound for scattering in $D$-dimensional spacetime
with one compact dimension has been derived. The upper bound for the
imaginary part of the hadronic scattering amplitude $T^D(s,t)$ was
found to be
\begin{equation}\label{bound_full}
\mathrm{Im}\,T_D(s,0) \leqslant s \,R_0^{D-2}(s)\,\Phi \!\left(
\frac{R_0}{R_c}, D \right).
\end{equation}
In \eqref{bound_full} the ``transverse radius'' is given by $R_0(s)
\sim t_0^{-1/2} \ln s$, where $t_0$ denotes the nearest singularity
in the $t$-channel. $R_c$ is the compactification radius of the ED,
and $\Phi(R_0/R_c,D)$ is a known function. At $R_c \gg R_0(s)$, the
inequality \eqref{bound_full} reproduces the Froissart-Martin bound
in a flat spacetime with arbitrary $D$ dimensions~\cite{Chaichian}
\begin{equation}\label{bound_large_scale}
\sigma_{\mathrm{tot}}^D(s) \leqslant \text{const}(D) \,R_0^{D-2}(s)
\;,
\end{equation}
while in the opposite limit $R_c \ll R_0(s)$ it results in the
inequality~\cite{Petrov:2001}
\begin{equation}\label{bound_small_scale}
\text{Im}\,T^D(s,0) \leqslant \text{const}(D)\,s \,R_0^{D-3}(s)\,
R_c \;.
\end{equation}

In \cite{Kisselev:2004} an analogous result has been obtained for
the scattering of two SM particles on a 3D brane embedded into a
flat spacetime with $n$ compact EDs ($D=4+n$). The inelastic cross
section $\sigma_{\mathrm{in}}^D(s)$ was calculated in the
transplanckian region $\sqrt{s} \gg M_D, |t|$, where $t$ is a
momentum transfer squared, and $M_D$ is a fundamental Planck scale
in $D$ dimensions. The result of the calculations is the following
\begin{equation}\label{bounds_cs}
\sigma_{\mathrm{inel}}^D(s) \simeq \text{const}(D) \times \left\{
\begin{array}{ll}
R_0^{2 + n}(s),    & R_c \gg \bar{R}_0(s) \;,  \vspace{.2cm} \\
R_0^2(s) \, R_c^n, & R_c \ll \bar{R}_0(s) \;,
\end{array}
\right.
\end{equation}
where $\bar{R}_0(s) = 2 R_g(s) \sqrt{\ln (s/M_D^2)}$, $R_g(s)$ being
the ``Regge gravitational radius'' (for more details, see
\cite{Kisselev:2004}).

To summarize, we can say that the dependence of physical observables
on the compactification radius of the ED(s) arises only when the
physical scale $R_{\mathrm{phys}}$ of the process becomes larger
than (comparable with) $R_c$. On the contrary, if $R_{\mathrm{phys}}
\ll R_c$, this dependence disappears (a physical process occurs on
distances $\sim R_{\mathrm{phys}}$, and it does not ``feel'' the
large scale $R_c$ at all).

\section{Conclusions} %

We have considered compactification of the asymptotically free
$\phi^3_{D=6}$ theory to manifolds $M_4 \otimes T^2/Z_2$ and
$M_4\otimes S^2/Z_2 $. The asymptotically free behavior of the
dimensionless triple coupling in $M_6$ is being inherited by
dimensional triple couplings of the light modes in both cases of
compactification, with details depending of the shape of
compactification. We also have considered the physical implications
for high-energy scattering which has the same energy dependence as
in simple four-dimensional case but retaining the compactification
radius as a parameter, when the interaction radius exceeds the
compactification size, while the ``memory'' of the latter disappears
at short-distance interactions which has now a different energy
dependence.



\setcounter{equation}{0}
\renewcommand{\theequation}{A.\arabic{equation}}

\section*{Appendix A} %

We give some useful properties of the two-dimensional inhomogeneous
Epstein zeta function $Z^a_2(s)$ \eqref{Epstein_func_def}, with
$a>0$. In \cite{Elizade:1998} the following explicit expression was
derived
\begin{equation}\label{Epstein_full}
Z^a_2(s) = - a^{-s} + \frac{\pi a^{1-s}}{s-1} +
\frac{A(s;a)}{\Gamma(s)} \;,
\end{equation}
where
\begin{align}\label{A}
A(s;a) = 4&\Bigg[ a^{1/4} \left( \frac{\pi}{\sqrt{a}}\right)^{\!\!s}
\sum_{n=1}^\infty n^{s-1/2} K_{s-1/2}(2\pi n\sqrt{a}) \nonumber \\
&+ a^{1/2} \left( \frac{\pi}{\sqrt{a}} \right)^{\!\!s} \sum_{n=1}^\infty n^{s-1}
K_{s-1}(2\pi n\sqrt{a}) \nonumber \\
&+ \sqrt{2} (2\pi)^s \sum_{n=1}^\infty n^{s-1/2} \sum_{d\|n}
d^{1-2s} \left( 2 + \frac{4a}{d^2} \right)^{\!1/4-s/2} \nonumber \\
&\times K_{s-1/2} \left( \pi n \sqrt{2 + \frac{4a}{d^2}} \right)
\Bigg] .
\end{align}
Here $K_\nu(z)$ is the modified Bessel function of the second kind,
$d\|n$ is the division of $n$. As one can see from \eqref{A},
$A(s;a)$ decreases exponentially, as $a \rightarrow \infty$. In the
limit $a \rightarrow 0$ the Chowla-Selberg formula
\cite{Chowla:1949} takes place
\begin{align}\label{Epstein_zero_a}
Z_2(s) &= 2 \zeta(2s) + 2\sqrt{\pi}
\,\frac{\Gamma(s-1/2)}{\Gamma(s)} \,\zeta(2s-1) \nonumber \\
&+ \frac{8\pi^s}{\Gamma(s)} \sum_{n=1}^\infty n^{s-1/2} \sum_{d\|n}
d^{1-2s} K_{s-1/2} (2\pi n) \;,
\end{align}
where
\begin{equation}\label{Z_2}
Z_2(s) \equiv Z^0_2(s)  = \sum_{n_1, n_2 \in
Z^{\!2}}{}^{\!\!\!\!\!'} \frac{1}{(n_1^2 + n_2^2)^s}
\end{equation}
is the two-dimensional Epstein zeta function \cite{Epstein:1903},
and $\zeta(s)$ is the Riemann zeta function \cite{Bateman_vol_1}.
Note that all (multi)series in \eqref{A}, \eqref{Epstein_zero_a} are
exponentially convergent.

The above formulas are valid over the \emph{whole} complex plane.
The inhomogeneous Epstein function \eqref{Epstein_full} exibits an
infinite number of simple poles at $s = 1, 1/2, -1/2, -3/2, \ldots$,
while the homogeneous Epstein function \eqref{Epstein_zero_a} has
only two simple poles at $s=1$ and $s=1/2$, with the residues $\pi$
and $-1/2$, respectively \cite{Kirsten:1991}. Note that both
functions are \emph{regular} at $s=0$. The formula
\begin{align}\label{Epstein_zero_eps}
Z^a_2(s)\big|_{s \rightarrow 0} &= -a^{-s} + \frac{\pi a^{1-s}}{s-1}
+ \frac{4}{\Gamma(s)} \sum_{n=1}^\infty \frac{1}{n} \bigg[
\frac{1}{2} \,e^{-2\pi n \sqrt{a}}
+ \sqrt{a} \,K_1(2\pi n \sqrt{a}) \nonumber \\
&+ \sum_{d\|n} d \,e^{-\pi n \sqrt{2 + 4a/d^2}} \bigg] +
\mathrm{O}(s^2)
\end{align}
gives an expansion of two-dimensional inhomogeneous Epstein function
around the point $s=0$.

The truncated Epstein-like zeta function is given by the expression
\cite{Elizade:1998}
\begin{align}\label{trancated_zeta_fun}
\zeta_t(s;a,q) &= \sum_{n=0}^\infty \frac{1}{[(n + a)^2 + q]^s} =
\left( \frac{1}{2} - a \right) \!q^{-s} +
\frac{q^{-s}}{\Gamma(s)}\sum_{m=1}^\infty
\frac{(-1)^m \Gamma(m+s)}{m!} \nonumber \\
&\times \zeta_H(-2m;a) \,q^{-m} + \frac{\sqrt{\pi} \,\Gamma(s -
1/2)}{2\Gamma(s)} \,q^{1/2-s}
\nonumber \\
&+ \frac{2\pi^s}{\Gamma(s)} \,q^{1/4-s/2} \sum_{n=1}^\infty
n^{s-1/2} \cos(2\pi n a) \,K_{s-1/2}(2\pi n \sqrt{q}) \;,
\end{align}
with $q > 0$. The first series on the right-hand side is asymptotic.
The last series decreases exponentially in parameter $q$. The
quantity
\begin{equation}\label{Hurwitw_fun}
\zeta_H(s;a) = \sum_{n=0}^\infty \frac{1}{(n + a)^s}
\end{equation}
is a Hurwitz zeta function \cite{Bateman_vol_1}. It is an analytic
function over the entire complex $s$-plane except the point $s=1$,
at which it has a simple pole. For $k=0, 1, 2, \ldots$, we have
\begin{equation}\label{Hurwitw_fun_values}
\zeta_H(-k;a) = -\frac{B_{k+1}(a)}{k+1} \;,
\end{equation}
where $B_r(a)$ is a Bernoulli polynomial \cite{Bateman_vol_1}. In
particular, $\zeta_H(0;a) = 1/2 - a$. In \eqref{trancated_zeta_fun},
apart form the pole at $s=1/2$, there is a whole sequence of poles
for $s = -1/2, -3/2, \ldots$.


%



\begin{thebibliography}{99}
%
\bibitem{Gross:1973}
D.J.~Gross and F.~Wilzek, \emph{Ultraviolet behavior of non-Abelian
gauge theories}, Phys. Rev. Lett. \textbf{30}, 1343 (1973);
\emph{Asymptotically free gauge theories. I}, Phys. Rev. D
\textbf{8}, 3633 (1973).
%
\bibitem{Politzer:1973}
H.D.~Politzer, \emph{Reliable perturbative results for strong
interactions?}, Phys. Rev. Lett. \textbf{30}, 1346 (1973).
%
\bibitem{Coleman:1973}
S.~Coleman and D.J. Gross, \emph{Price of asymptotic freedom}, Phys.
Rev. Lett. \textbf{31}, 851 (1973).
%
\bibitem{Bond:2017}
A.D.~Bond and D.F.~Litim, \emph{Theorems for asymptotic safety of
gauge theories}, Eur. Phys. J. C \textbf{77}, 429 (2017);
\emph{Price of asymptotic safety}, Phys. Rev. Lett. \textbf{122},
211601 (2019).
%
\bibitem{Gross:1974}
D.J. Gross and A.~Neveu, \emph{Dynamical symmetry breaking in
asymptotically free field theories}, Phys. Rev. D \textbf{10}, 3235
(1974).
%
\bibitem{Brezin:1976}
E.~Brezin and J.~Zinn-Justin, \emph{Renormalization of the nonlinear
$\sigma$ model in $2+\varepsilon$ dimensions —- application to the
Heisenberg ferromagnets}, Phys. Rev. Lett. \textbf{36}, 691 (1976);
E.~Brezin, J.~Zinn-Justin and L.Le Guilou, \emph{Renormalization of
the nonlinear $\sigma$ model in $2+\varepsilon$ dimensions}, Phys.
Rev. D \textbf{14}, 2615 (1976).
%
\bibitem{Symanzik:1973}
K.~Symanzik, \emph{A Field Theory with Computable Large-Momenta
Behaviour}, Lett. Nuovo Cim. \textbf{6}, 77 (1973).
%
\bibitem{Brandt:1976}
R.A.~Brandt, \emph{Asymptotically free $\phi^4$ theory}, Phys. Rev.
D \textbf{4}, 3381 (1976).
%
\bibitem{Macfarlane:1974}
A.J.~Macfarlane and G.~Woo, \emph{$\phi^3$ theory in six dimensions
and the renormalization group}, Nucl. Phys. B \textbf{77}, 91
(1974); [Erratum: Nucl. Phys. B \textbf{86}, 548 (1975)].
%
\bibitem{Elizade:1994}
E.~Elizade and Yu.~Kubyshin, \emph{Possible evidence of Kaluza-Klein
particles in a scalar model with spherical compactification}, J.
Phys. A: Math. Gen. \textbf{27}, 7533 (1994).
%
\bibitem{Lopez:2020}
V.A.~L\'{o}pez-Ozorio \emph{et al.}, \emph{One-loop order effects
from one universal extra dimension on $\lambda \phi^4$ theory},
arXiv:2002.10015.
%
\bibitem{Akhoury:2009}
R.~Akhoury and C.S.~Gauthier, \emph{Decoupling of heavy Kaluza-Klein
modes in models with five-dimensional scalar fields}, J. Phys. G:
Nucl. Part. Phys. \textbf{36}, 015005 (2009).
%
\bibitem{Akhoury:2008}
R.~Akhoury and C.S.~Gauthier, \emph{Kaluza-Klein model with
spontaneous symmetry breaking: Light-particle effective action and
its compactification scale dependence}, Phys. Rev. D \textbf{78},
105002 (2008).
%
\bibitem{Martinez:2020}
E.~Matrinez-Pascual \emph{et al.}, \emph{Implications of extra
dimensions in the effective charge and the beta function in quantum
electrodynamics}, Phys. Rev. D \textbf{101}, 035034 (2020).
%
\bibitem{Sochichiu:1999}
C.~Sochichiu, \emph{Quantum Kaluza-Klein compactification}, Phys.
Lett. B \textbf{463}, 27 (1999).
%
\bibitem{Sochichiu:2000}
C.~Sochichiu, \emph{Compactified quantum fields. Is there life
beyond the cut-off scale?}, Phys. Lett. B \textbf{477}, 253 (2000).
%
\bibitem{Dohi:2010}
H.~Dohi and K.~Oda, \emph{Universal extra dimensions on real
projective plane}, Phys. Lett. B \textbf{692}, 114 (2010).
%
\bibitem{Maru:2010}
N.~Maru, T.~Nomura, J.~Sato and M.~Yamanaka, \emph{The universal
extra dimensional model with $S^2/Z_2$ extra-space}, Nucl. Phys. B
\textbf{830}, 414 (2010).
%
\bibitem{Maru:2014}
N.~Maru, T.~Nomura and J.~Sato, \emph{One-loop radiative correction
to Kaluza-Klein mases in $S^2/Z_2$ universal extra dimensional
model}, Prog. Theor. Phys. 083B04 (2014).
%
\bibitem{Dohi:2014}
H.~Dohi \emph{et al.}, \emph{Notes on sphere-based universal extra
dimensions}, Afr. Rev. Phys. \textbf{9}, 0066 (2014).
%
\bibitem{Kakuda:2013}
T.~Kakuda, K.~Nishiwaki and R.~Watanabe, \emph{Universal extra
dimensions after Higgs discovery}, Phys. Rev. D \textbf{88}, 035007
(2013).
%
\bibitem{Collins}
J.C.~Collins, \emph{Renormalization}, Cambridge University Press,
Cambridge, 1984.
%
\bibitem{Kosyakov:2001}
B.P.~Kosyakov, \emph{Physical sense of renormalizability}, Phys.
Part. Nucl. \textbf{32}, 488 (2001).
%
\bibitem{t'Hooft:1973}
G.~t'Hooft, \emph{Dimensional regularization and the renormalization
group}, Nucl. Phys. B \textbf{61}, 455 (1973).
%
\bibitem{Yndurain}
F.J.~Yndurain, \emph{Quantum Chromodynamics. An Introduction to the
Theory of Quarks and Gluons}, Springer-Verlag, 1983.
%
\bibitem{Kompaniets:2021}
M.~Kompaniets and A.~Pikelner, \emph{Critical exponents from
five-loop scalar theory renormalization near six-dimensions},
arXiv:2101.10018.
%
\bibitem{Vasiliev}
A.N.~Vasiliev, \emph{The field theoretic renormalization group in
critical behavior theory and stochastic dynamics}, CRC Press
Company, NY, 2020.
%
\bibitem{Kirsten:1991}
K.~Kirsten, \emph{Inhomogeneous multidimensional Epstein zeta
function}, J. Math. Phys. \textbf{32}, 3008 (1991);
\emph{Generalized multidimensional Epstein zeta function}, J. Math.
Phys. \textbf{35}, 458 (1994).
%
\bibitem{Dowker:1976}
J.S.~Dowker and R. Critchley, \emph{Effective Lagrangian and
energy-momentum tensor in de Sitter space}, Phys. Rev. D
\textbf{13}, 3224 (1976).
%
\bibitem{Hawking:1997}
S.W. Hawking, \emph{Zeta-function regularization of path integrals
in curved space-time}, Comm. Math. Phys. \textbf{55}, 133 (1977).
%
\bibitem{Zwiebach}
B.~Zwiebach, \emph{A first cours in string theory}, 2nd edition,
Cambridge University Press, UK, 2009.
%
\bibitem{Bateman_vol_1}
\emph{Higher Transcendental Functions}. Vol.~1. By the staff of the
Bateman manuscript project (A.~Erd\'{e}lyi,\emph{ Editor};
W.~Magnus, F.~Oberhettinger, F.G.~Tricomi, \emph{Associates}),
McGraw-Hill Book Company, New York, 1953.
%
\bibitem{Shirkov:1997}
D.V.~Shirkov and I.L.~Solovtsov, \emph{Analytic model for the QCD
running coupling with universal $\bar{\alpha}_s(0)$ value}, Phys.
Rev. Lett. \textbf{79}, 1209 (1997); \emph{Ten years of the analytic
perturbation theory in QCD}, Theor. Math. Phys. \textbf{150}, 132
(2007).
%
\bibitem{Duff:1981}
M.J.~Duff and D.J.~Toms, \emph{Divergences and anomalies in
Kaluza-Klein theories}, CERN-3248. Proceedings of the 2nd Seminar on
Quantum Gravity, edited by M.A.~Markov and P.C.~West, Moscow, USSR,
13-15 October 1981, pp.431-461.
%
\bibitem{MacRobert}
T.M.~MacRobert, \emph{Spherical Harmonics: An Elementary Treatise on
Harmonic Functions with Applications}, 3rd edition, Pergamon Press,
1967.
%
\bibitem{Petrov:2001}
V.A.~Petrov, \emph{Froissart-Martin bound in spaces with compact
extra dimensions}, Mod. Phys. Lett. A \textbf{16}, 151 (2001);
\emph{Froissart-Martin bound in spaces of compact dimensions}, Phys.
of Atom. Nucl. \textbf{65}, 877 (2002); \emph{High-energy behaviour
in four and more dimensions}, in Proceedings of the International
Conference on Theoretical Physics (TH 2002), 22-26 July 2002, Paris,
France, pp. 253-255.
%
\bibitem{Chaichian}
M.~Chaichian and J.~Fisher, \emph{Unitarity bounds for high-energy
scattering in many dimensions}, in Proceedings of the International
Conference ``Hadron Structure '87'', Smolenice, CSSR, November
16-20, 1987, p.~334; \emph{Higher-dimensional space-time and unitary
bound on the scattering amplitude}, Nucl. Phys. B \textbf{303}, 557
(1988).
%
\bibitem{Kisselev:2004}
A.V.~Kisselev and V.A.~Petrov, \emph{Gravireggeons in extra
dimensions and interaction of ultra-high energy cosmic neutrinos
with nucleons}, Eur. Phys. J. C. \textbf{36}, 103 (2004).
%
\bibitem{Elizade:1998}
E.~Elizade, \emph{Multidimentional Extension of the generalized
Chowla-Selberg formula}, Comm. Math. Phys. \textbf{198}, 83 (1998);
\emph{Zeta functions: formulas and applications}, Comp. Appl. Math.
\textbf{118}, 125 (2000).
%
\bibitem{Chowla:1949}
S.~Chowla and A.~Selbergs, \emph{On Epstein's zeta function}, Proc.
Nat. Acad. Sci. U.S.A. \textbf{35}, 317 (1949).
%
\bibitem{Epstein:1903}
P.~Epstein. \emph{Zur Theorie allgemeiner Zetafunctionen}, Math.
Ann., \textbf{56}, 615 (1903), \emph{Zur Theorie allgemeiner
Zetafunctionen. II}, Math. Ann., \textbf{63}, 205 (1906).
%
\end{thebibliography}
\end{document}